\pdfoutput=1
\documentclass[9pt,conference]{IEEEtran}

\usepackage{amsmath}
\usepackage{amssymb}
\usepackage{graphicx}
\usepackage{float}
\usepackage{array}
\usepackage{tabulary}

\begin{document}
\title{Inference of an explanatory variable from observations in a high-dimensional space: Application to high-resolution spectra of stars}

\author{
\IEEEauthorblockN{{\large Victor Watson$^{1,2}$, Jean-Fran\c{c}ois Trouilhet$^{1,2}$ (IEEE senior member), Fr\'ed\'eric Paletou$^{1,2}$ and St\'ephane Girard$^3$}}
\vspace{0.5em}
\IEEEauthorblockA{(1) Universit\'e de Toulouse, UPS-Observatoire Midi-Pyr\'en\'ees, Irap, Toulouse, France \\ (2) CNRS,  Institut de Recherche en Astrophysique et Plan\'etologie, 14 av. E. Belin, 31400 Toulouse, France \\
(3) INRIA, 655 Avenue de l'Europe, 38330 Montbonnot-Saint-Martin, France \\ 
Email : \{victor.watson, jean-francois.trouilhet, frederic.paletou\}@irap.omp.eu , stephane.girard@inria.fr}
}

\maketitle

\begin{abstract}
Our aim is to evaluate fundamental parameters from the analysis of the electromagnetic spectra of stars.
We may use $10^3$-$10^5$ spectra; each spectrum being a vector with $10^2$-$10^4$ coordinates.
We thus face the so-called ``curse of dimensionality". We look for a method to reduce the size of this data-space, keeping only the most relevant information.
As a reference method, we use principal component analysis (PCA) to reduce dimensionality. However, PCA is an unsupervised method, therefore its subspace was not consistent with the parameter. We thus tested a supervised method based on Sliced Inverse Regression (SIR), which provides a subspace consistent with the parameter. It also shares analogies with factorial discriminant analysis: the method slices the database along the parameter variation, and builds the subspace which maximizes the inter-slice variance, while standardizing the total projected variance of the data. Nevertheless the performances of SIR were not satisfying in standard usage, because of the non-monotonicity of the unknown function linking the data to the parameter and because of the noise propagation. We show that better performances can be achieved by selecting the most relevant directions for parameter inference.

Preliminary tests are performed on synthetic pseudo-line profiles plus noise.
Using one direction, we show that compared to PCA, the error associated with SIR is 50$\%$ smaller on a non-linear parameter, and 70$\%$ smaler on a linear parameter. Moreover, using a selected direction, the error is 80$\%$ smaller for a non-linear parameter, and 95$\%$ smaller for a linear parameter. 
\end{abstract}

\IEEEpeerreviewmaketitle

\section{Introduction}\label{seq_intro}

As an historically data-intensive science, astrophysics has obviously to deal with significant data-processing issues. Projects such as the Gaia 
 survey \cite{gaia2}
 or LSST\footnote{https://www.lsst.org/} 
are acquiring and will acquire a lot of data, both in terms of number of samples and in terms of amount of data by sample.
We are interested in methods able to extract as much physical content as possible from such large volumes of data.

Our problem deals with high dimensional data-processing.
Similarly to \cite{Paletou15}, our goal is to infer stellar fundamental parameters such as effective temperature, surface gravity or metallicity. We also want to infer an important physical parameter which may significantly affect the appearance of stellar spectra, the projected rotational velocity.

These physical parameters values are to be estimated from the stars measured spectra. Nowadays precision optical spectroscopy relies on observations made over large spectral bandwidths, of the order of about 500 nm or more (without any gap), and at high-spectral resolution, typically of several $10^4$ in resolvance (resolvance is related to sampling properties of these instruments $R = \frac{\lambda}{\Delta\lambda}$).

For these problems, there is no direct linear relationship between the data and the physical parameters. To ease the search for such a function or relationship, we use methods based on a reduction of the dimensionality of our data-space. 
These methods should extract from the data the information related to the physical parameter we want to infer.
One of the most popular among these methods is PCA \cite{Jolliffe} which is the method used by \cite{Paletou15}. 
In this study, our aim is to find the best subspace for the estimation of an explanatory variable $y$ from a data vector $x$. This subspace is to be estimated from a database containing many examples of pairs ($x_i$,$y_i$). Once the subspace is found, a k-Nearest-Neighbours method will give the estimator $\hat{y}_i$ of the unknown value $y_i$ associated with the data vector $x_i$.
Hereafter we will propose an alternative solution based on SIR \cite{li91} to improve the results. Unlike PCA, SIR is a supervised method. Hence, when building the projection subspace, we will be able to add the information about the physical parameters (or explanatory variables) we want to retrieve from the data. These parameters are considered independently one at a time. SIR method has already been used in \cite{bernard-michel}. One of the differences with our case is that we have no information about the dimension of the subspace consistent with the one-dimensional manifold described by the physical parameter; and we cannot use a physical modelling to determine the dimension of this subspace.

SIR suffers from a limitation due to the ill-conditioning of the variance-covariance matrix. Thus, it will sometimes require some preprocessing in order to perform a stable inversion of the variance-covariance matrix. This preprocessing will consist in applying a PCA on the data before applying SIR.

The first section briefly recalls the basic principles of PCA and SIR methods. The second section presents the implementation of SIR and some examples to enlighten how it basically works. The third section is about testing SIR method on a simplified data set which mimics the astrophysical data we want to deal with. The fourth section discusses the tuning of the method and the pre-processing.

\section{Methods}
\subsection{Principal Component Analysis}
PCA is a method which aims at finding the subspace that preserves the most the variance of the projected data.

Considering a data matrix $\textbf{X}$ of dimension $\textit{M} \times \textit{N}$ with each line being an element from the database, PCA finds the projection matrix $\textbf{V}$ such as $\textbf{X}^T\textbf{V}$ has the maximum variance. The first vector $v_1$ is the one that maximizes $Var(\textbf{X}^Tv_1)$ under the constraint $v_1^Tv_1 = 1$, the second vector $v_2$ is the vector orthogonal to $v_1$ that maximizes $Var(\textbf{X}^Tv_2)$ under the same constraint, and so on until the number of components reaches the specified dimension of the projection subspace.
To find these vectors, one can perform an eigen-decomposition of the variance-covariance matrix of $\textbf{X}$ denoted by $\boldsymbol{\Sigma}$. Thus we will find these vectors by solving: $(\boldsymbol{\Sigma} - \lambda \textbf{I}_M)v = 0$ \cite{Jolliffe}.

The first step of PCA is to compute the variance-covariance matrix $\boldsymbol{ \Sigma }$:

\begin{equation}\label{eq_sigma}
\boldsymbol{\Sigma}_{i,j} = \frac{1}{M}\sum_{k=1}^{M} (\textbf{X}_{k,i} - \overline{\textbf{X}_i})  (\textbf{X}_{k,j} - \overline{\textbf{X}_j} )
\end{equation}
where $N$ is the dimension of the original data-space, $\textbf{X}_k$ is the $k^{th}$ column containing all the $x$ vectors projected on the $k^{th}$ component of the original data-space. $\overline{\textbf{X}_i}$ and $\overline{\textbf{X}_j}$ are the mean values computed on lines $i$ and $j$.

The subspace obtained by PCA is spanned by the eigenvectors of $\boldsymbol{\Sigma}$ associated to its greatest eigenvalues.

Let $\textbf{V}$ be the $N \times P$ matrix with every column being one of these $P$ selected eigenvectors, and $N$ being the dimension of the original space. The $x$ vectors can be projected on the subspace associated with PCA:
\begin{equation}
x_p = (x^T - \overline{\textbf{X}}) \textbf{V}
\end{equation} 
where $x$ is the representation of an element from the database in the original data-space, and $\overline{\textbf{X}}$ is the mean of the elements from the database in the original data-space. $\overline{\textbf{X}}$ is computed on all the lines of the matrix $\textbf{X}$.

The main drawback of PCA on our application is that this method does not take into account the prior knowledge on the explanatory variable we want to infer, as it is an unsupervised method. This is why we decided to investigate SIR as an alternative.

\subsection{Sliced Inverse Regression}

SIR \cite{li91} aims at finding the directions that explain the best the variation of the explanatory variable $y$ (taken one by one) from the data $x$. The principle of the method is very close to the one of the PCA: it relies on the estimation of a linear projection subspace. However SIR takes into account the information on the variation of the explanatory variable in the building of this subspace. The database is divided into $H$ slices along the variation of the explanatory variable, each slice being composed of pairs ($x$,$y$) which have close values of the explanatory variable. SIR builds a subspace that maximizes the variance between the slices while minimizing the variance within the slices. Somehow, SIR can be compared to a Linear Discriminant Analysis (LDA)\cite{LDA} in a classification problem. Thus, SIR finds a linear projection that drives apart the elements that have very different values of the explanatory variable while bringing together the elements which have close values of the explanatory variable.
Let us denote by $\textbf{X}_h$ the elements from the database in the slice $h$. For every slice we compute the mean $\overline{m}_h$ 
\begin{equation}
\overline{m}_h = \frac{1}{n_h} \sum_{\textbf{X}_i \in h}  \textbf{X}_i 
\label{eq_mh}
\end{equation}
where $n_h$ is the number of elements in the slice $h$,
so that we have the matrix $\textbf{X}_H$ with every line being the mean $\overline{m}_h$ associated with the elements of the slice $h$. We then compute the variance-covariance of the slices means:
\begin{equation}\label{eq_gamma}
\boldsymbol{\Gamma} = \frac{1}{H}(\textbf{X}_H - \overline{\textbf{X}_H})^T(\textbf{X}_H -  \overline{\textbf{X}_H}).
\end{equation} 

The directions computed by SIR will be given by the eigenvectors associated with the greatest eigenvalues of $\boldsymbol{\Sigma}^{-1}\boldsymbol{\Gamma}$. The vectors spanning the sub-space we are looking for are the solutions of  $(\boldsymbol{\Sigma}^{-1}\boldsymbol{\Gamma} - \lambda \textbf{I}_M)v = 0$.
\vspace{-0.2cm}
\section{Sliced Inverse Regression}

\subsection{Implementation}

SIR can be computed the following way:

\begin{itemize}
\item compute the variance-covariance matrix $\boldsymbol{\Sigma}$ as in eq. \ref{eq_sigma}
\vspace{-0.2cm}
\item sort the parameter vector $y$ so it becomes $y_{sorted}$
\vspace{-0.2cm}
\item split the sorted vector $y_{sorted}$ in $H$ non-overlapping slices
\vspace{-0.2cm}
\item for each slice $h$ compute the mean $\overline{m}_h$ according to eq. \ref{eq_mh}
\vspace{-0.2cm}
\item set the $\boldsymbol{\Gamma}$ matrix as in eq. \ref{eq_gamma}
\vspace{-0.2cm}
\item build a projection matrix $\textbf{V}$ so that the columns of $\textbf{V}$ are the eigenvectors associated with the greatest eigenvalues of $\boldsymbol{\Sigma}^{-1}\boldsymbol{\Gamma}$.

\end{itemize}  

One of the critical point of this method may be the inversion of $\boldsymbol{\Sigma}$. Indeed, $\boldsymbol{\Sigma}$ can be ill-conditioned as discussed in section~4. In that case, we propose to apply a PCA to project the data on a subspace that maximizes the variance. Doing so, PCA will lead to an inversion of $\boldsymbol{\Sigma}$ more stable, and in the meantime, it will reduce the dimension of the dataspace and thus requires less  computational resources. There is a compromise to find between improving enough the conditioning and loosing information held by the data. This will be discussed in section 5.

\subsection{Examples}

\subsubsection{Linear example}\label{sec_linex}

As a first example we will consider a linear relationship between $x$ and $y$ such as $y = \beta^Tx +\epsilon$. The dimension of $x$ is 4 and the database contains 1000 elements. $\textbf{X}$ follows a normal standard distribution $\textbf{X} \sim N(\mu,I_4)$, where $\mu$ is a null vector of dimension $\textit{4} \times \textit{1}$ and $I_4$  is a dimension 4 identity matrix, $\beta^T = [1,1,0,1]$ and the noise $\epsilon \sim N(0,0.1 I_4)$.

\begin{figure}[H]
\centering
\includegraphics[height = 4cm,width =8cm]{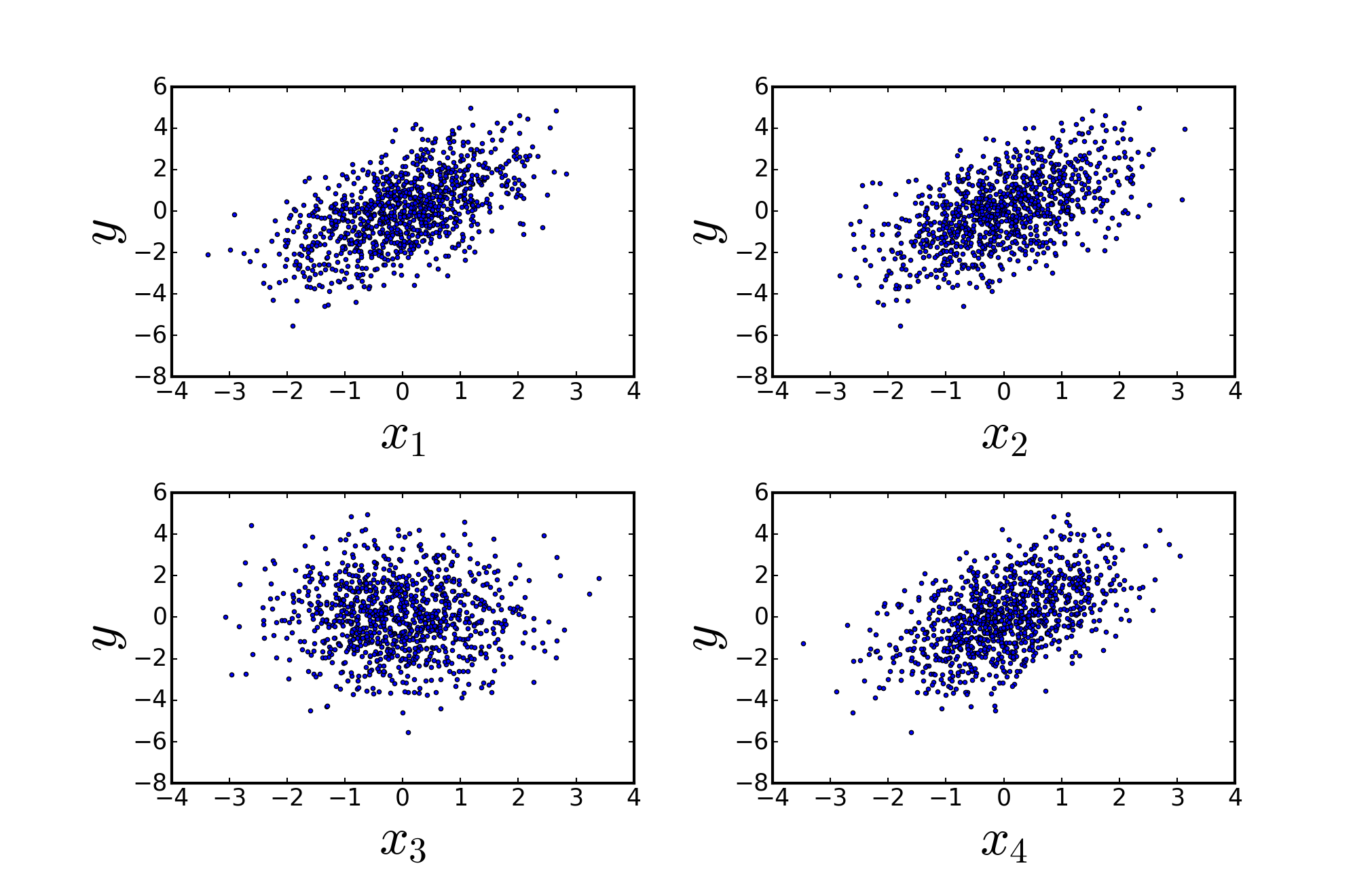}
\caption{Values of $y$ versus the realisations of $\textbf{X}$ for the example of section~\ref{sec_linex}. We can observe that none of these directions $x_i$ of the original data-space is relevant to explain the variations of the explanatory variable $y$. This can be explained by the $x_i$ varying independently from one to another, $y$ being a linear combination of these $x_i$, and all these components having the same weight.}
\label{ex1}
\end{figure}

\vspace{-0.6cm}
Fig. \ref{ex1} shows that from every component of $\textbf{X}$, it is impossible to determine precisely $y$. It appears in these four sub-figures that for any value of the $x_i$, $y$ looks like it is randomly distributed. This is why one must find the correct linear combination of the $x_i$ to explain $y$.

\begin{figure}[H]
\centering
\includegraphics[height = 4cm,width =8cm]{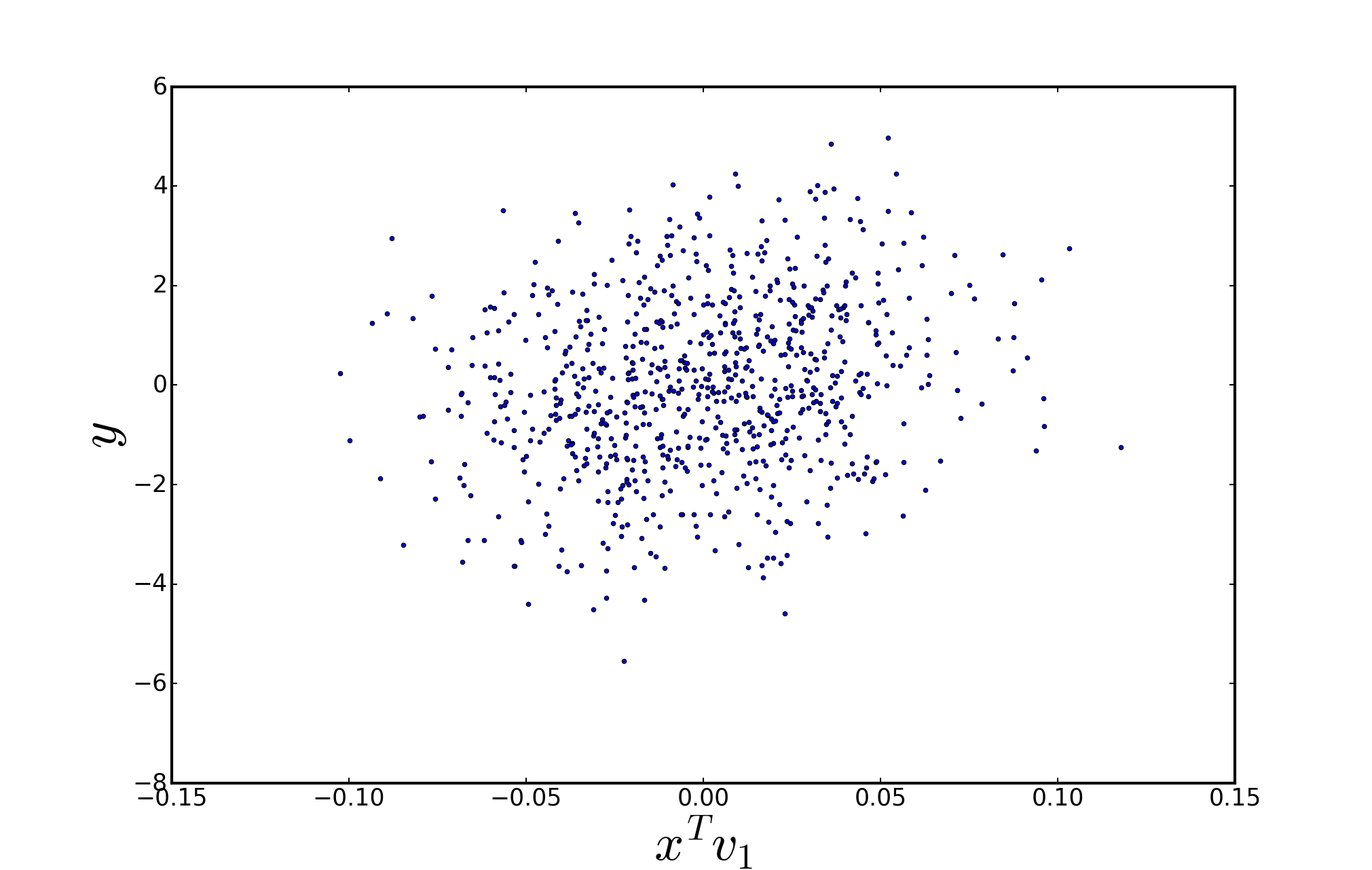}
\caption{Values of $y$ versus the realisations of $\textbf{X}$ projected on the first direction given by PCA for the example of section \ref{sec_linex}. Such a cloud of points indicates that the variance is nearly irrelevant to infer the value of $y$ from a known value of $x$.}
\label{PCA_ex1}
\end{figure}

The first direction $v_1$ given by PCA, displayed in Fig. \ref{PCA_ex1}, is the one that maximizes the variance in the data, and it appears that this criterion is not relevant here, because as in Fig. \ref{ex1}, $y$ seems to be randomly distributed for any value of the projected data $x^Tv_1$.

\begin{figure}[H]
\centering
\includegraphics[height = 4cm,width =8cm]{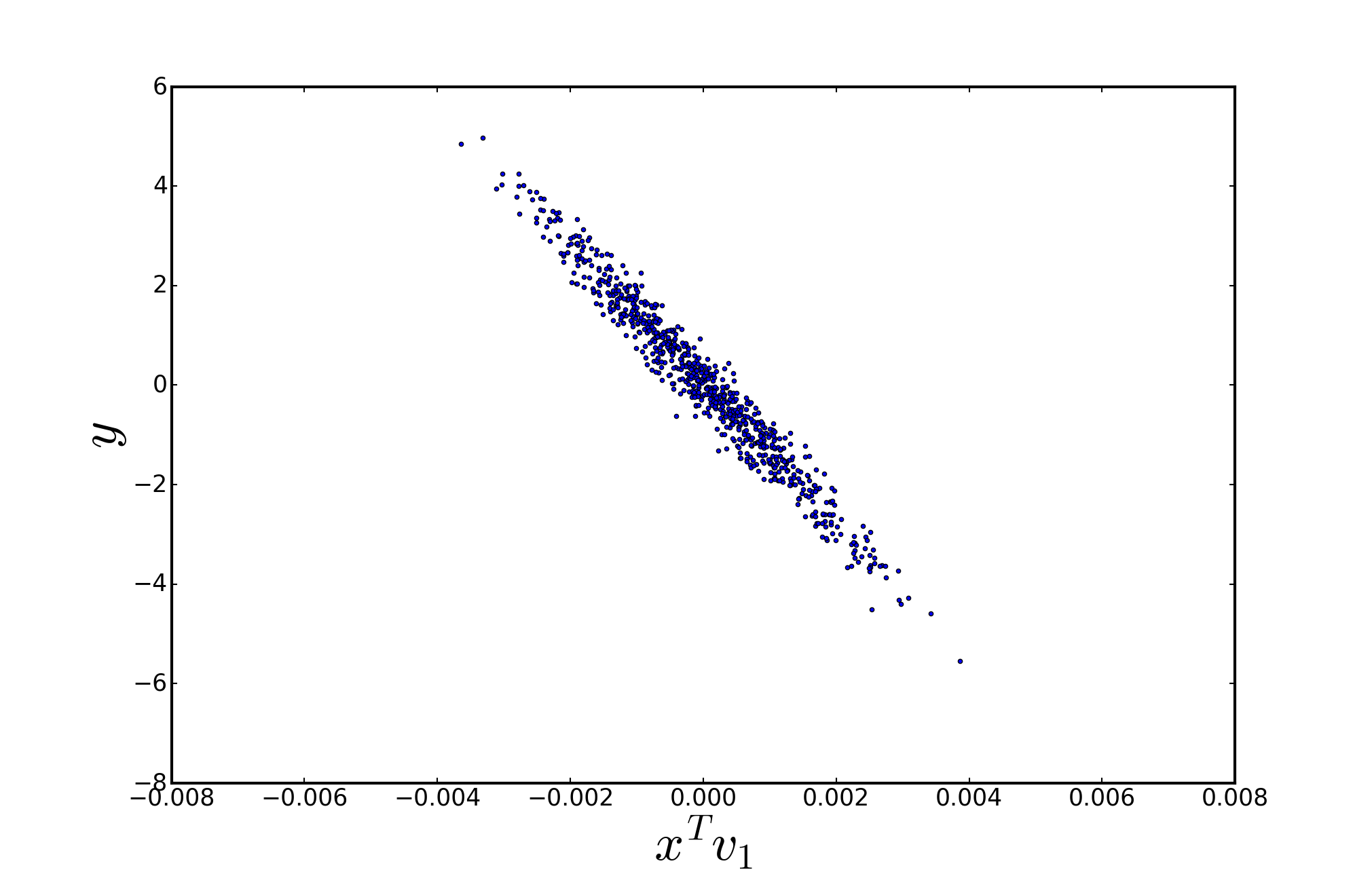}
\caption{Values of $y$ versus the realisations of $\textbf{X}$ projected on the first direction given by SIR for the example of section \ref{sec_linex}. This figure clearly shows that SIR is able to find the linear projection which explains $y$ from the components presented on Fig.\ref{ex1}.}
\label{SIR_ex1}
\end{figure}

Fig. \ref{SIR_ex1} shows that SIR provides a direction $v_1$ that allows us to determine quite precisely $y$ from $x^Tv_1$. Indeed for every value of $x^Tv_1$ the range of values that $y$ can take is limited.

This first example shows that the first direction given by SIR (Fig. \ref{SIR_ex1}) finds a much more relevant combination of $\textbf{X}$ to explain $y$ than PCA (Fig. \ref{PCA_ex1}) even though it is quite difficult to infer that relationship from $\textbf{X}$ in the original space (Fig. \ref{ex1}). In this case, the condition number associated with $\boldsymbol{\Sigma}$ is 1.26, so there is no need for any preprocessing.

\subsubsection{Nonlinear example}\label{sec_nonlinex}

The second example focusses on a non linear case. All the components of $\textbf{X}$ are independent and follow a standard uniform distribution $U[0,1]$ and $y(x) = 2x_1x_2+x_3+0  x_4 + \epsilon$, where the noise is $\epsilon \sim N(0,0.01 I_4)$.

\begin{figure}[H]
\centering
\includegraphics[height = 4cm,width =8cm]{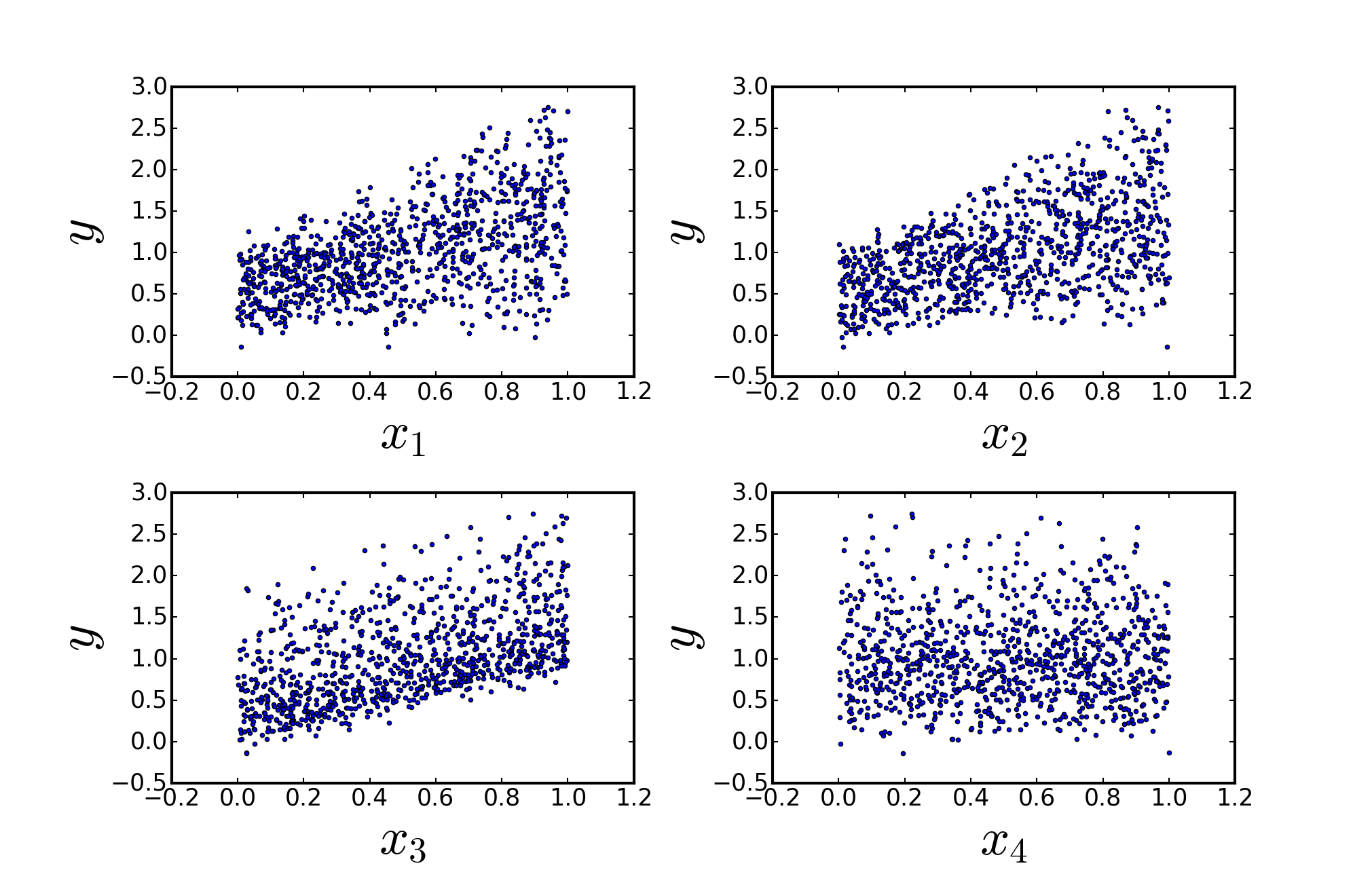}
\caption{Values of $y$ versus the realisations of $\textbf{X}$ for the example of section \ref{sec_nonlinex}. As in the case of the linear example, it is not obvious to find a relationship between the data in its original space and the explanatory variable even if we can see that the three first components (top-left, top-right, and bottom-left) of the data-space hold information about $y$. This is due to the fact that the range of the values taken by $y$ is rather wide whatever the value of $x_i$ is.}
\label{ex2}
\end{figure}

Once again in this example, we have to find a good combination of the components $x_i$ shown on Fig. \ref{ex2} to explain $y$.

\begin{figure}[H]
\centering
\includegraphics[height = 4cm,width =8cm]{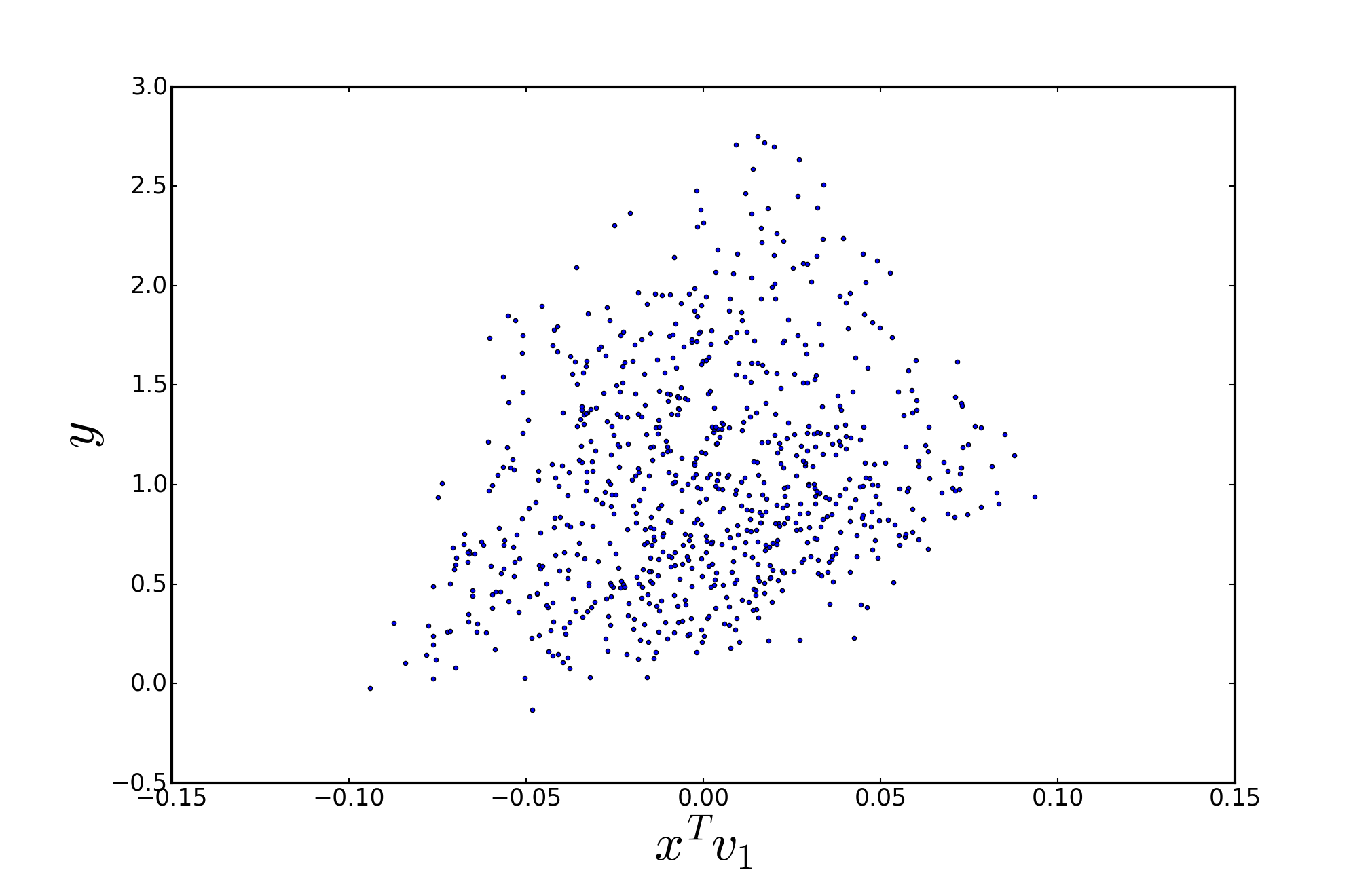}
\caption{Values of $y$ versus the realisations of $\textbf{X}$ projected on the first directions given by PCA for the example of section \ref{sec_nonlinex}. It appears that the first direction given by PCA holds almost no information about the explanatory variable because no matter what $x^Tv_1$ is, $y$ cannot be precisely determined.}
\label{PCA_ex2}
\end{figure}

We show with Fig. \ref{PCA_ex2} that, using PCA, a relationship appears between $x^Tv_1$ and $y$. Indeed, the scatter-plot (Fig. \ref{PCA_ex2}) exhibits a shape and is not fully random. Nevertheless, this so-called 'shape' is far from being sufficiently narrow to determine precisely $y$ from $x^Tv_1$.

\begin{figure}[H]
\centering
\includegraphics[height = 4cm,width =8cm]{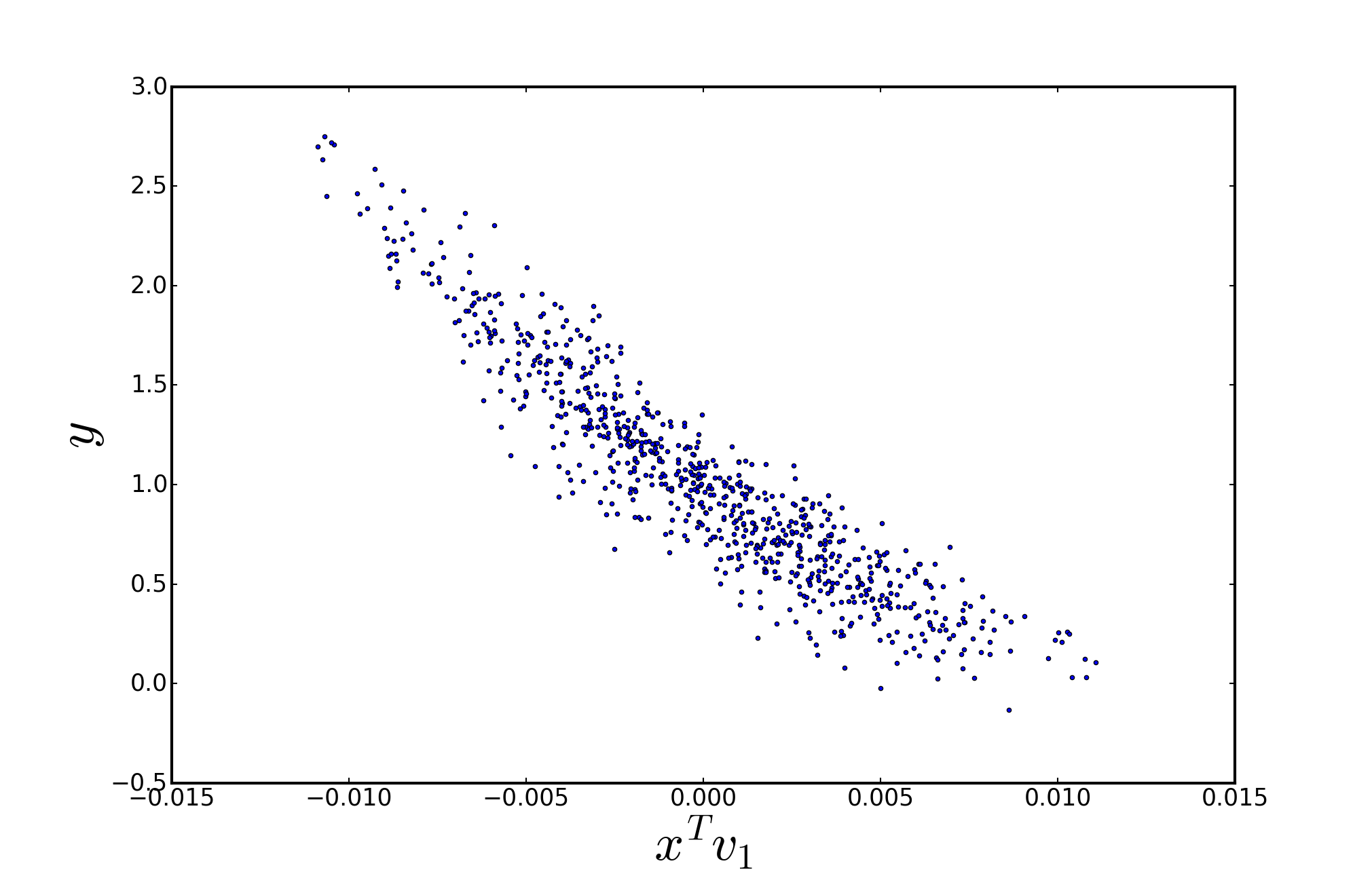}
\caption{Values of $y$ versus the realisations of $\textbf{X}$ projected on the first direction given by SIR for the example of section \ref{sec_nonlinex}. Even though the direction is not as efficient as in the linear case, we can see here that SIR projects the data on a direction which highly correlates the data with the explanatory variable.}
\label{SIR_ex2}
\end{figure}

Again in the case where the relationship between the data and the parameter is not linear, SIR (Fig. \ref{SIR_ex2}) provides more relevant directions to explain $y$ from $\textbf{X}$ than PCA. Of course, since it is a linear projection method more than one direction can be necessary to estimate correctly the one dimensional explanatory variable which is non-linearly linked to the data. In this example, a preprocessing of the data was not necessary because the condition number associated with $\boldsymbol{\Sigma}$ is 1.17.

\section{Application to synthetic data}
\subsection{Inference of simple Gaussian lines parameters}

We showed that SIR can determine the combinations of $\textbf{X}$ that best explain $y$. We will try to use it to obtain $\hat{y}$ when it is unknown.
Hereafter, the database is a set of simplified Gaussian lines that mimic the astrophysical data of interest. It is a set of $10^3$ Gaussian $\textbf{X}$ in which 800 will be used as a reference ($y$ is known) database $\textbf{X}_0$ and the 200 others will be used as a test database $\textbf{X}_i$ ($y$ is to be determined). Each Gaussian $x_j$ is associated with a set of explanatory variables $h_j$ and $w_j$, respectively the depression and the width of a simplified Gaussian spectral line: 
\begin{equation}\label{eq_gauss}
x(\lambda;h,w) = 1 - h \times e^{-\frac{1}{2}\left(\frac{\lambda-\lambda_0}{w}\right)^2} + \epsilon .
\end{equation}
$h$ and $w$ will be processed independently, 
$\lambda_0=0$ is common to all the elements. $h$ and $w$ are uniformly distributed $h \sim U[0.1,0.9]$, $w \sim U[1,3]$ and the noise $\epsilon$ is normally distributed $\epsilon \sim N(0,\sigma_\epsilon)$. 

For this example we will add a noise $\epsilon$ such as $\sigma_\epsilon = 0.005$ equivalent to a signal to noise ratio around 32dB. 

Once the data is projected on the subspace computed by SIR we will proceed with a simple mean on a 10 nearest neighbours method, to infer the value of the explanatory variable $\hat{y}_i$ of an individual $x_i$:

\begin{equation}
\hat{y}_i = \frac{1}{10} \sum_{y_{0_j} \in W_i}y_{0_j}.
\end{equation}

Here, $W_i$ is the set of the values of the explanatory variable from the training data-base $x_{p}$, associated with the 10 nearest neighbours of $x_{p_i}$. 

\begin{figure}[H]
\centering
\includegraphics[height = 4.5cm,width =8cm]{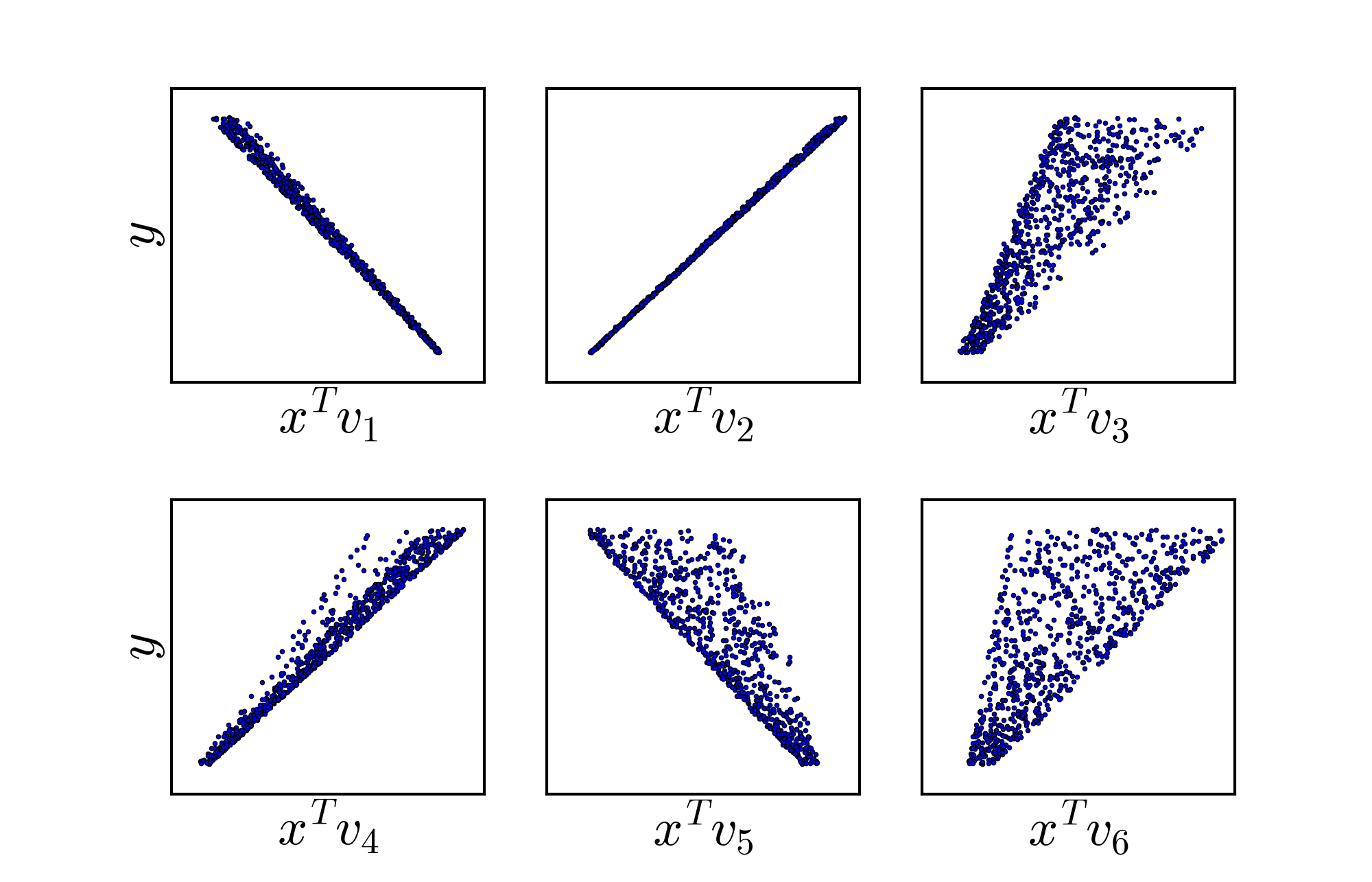}
\caption{Values of the depressions of the Gaussians $y$ versus the realisations of $\textbf{X}_0$ projected on the 6 directions given by SIR. Most of the information we seek is held by the first two directions (top-left and top-center), but in some cases the directions 3 (top-right), 4 (bottom-left) or 5 (bottom-center) can add quite precise information about $y$.}
\label{SIR_dep_dir}
\end{figure}
\begin{figure}[H]
\centering
\includegraphics[height = 4.5cm,width =8cm]{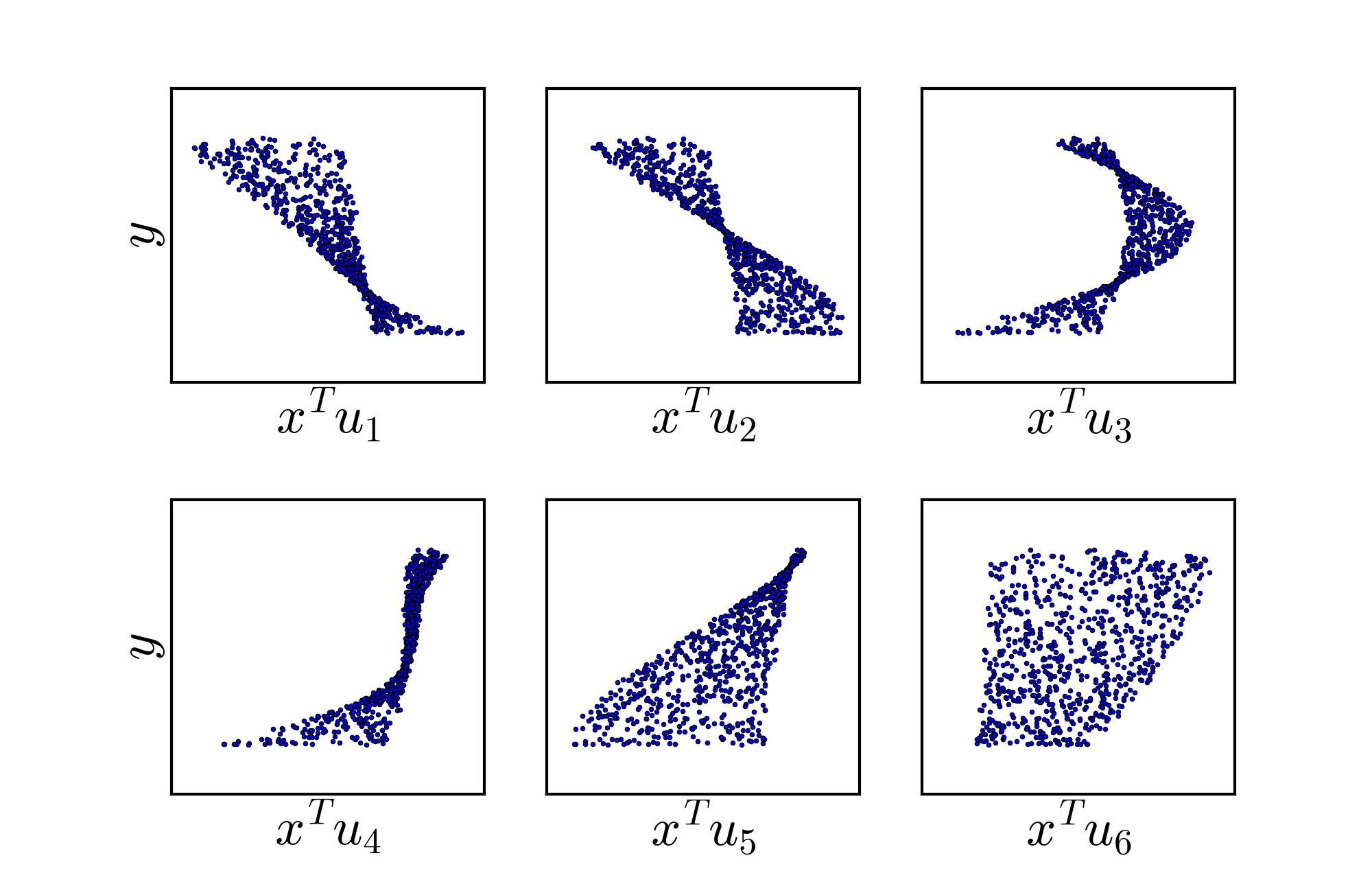}
\caption{Values of the widths of the Gaussians $y$ versus the realisations of $\textbf{X}_0$ projected on the 6 directions given by SIR. In that case the link between the explanatory variable and the data is non-linear so it does not seem to be a subspace which explains correctly all the values that can be taken by $y$.}
\label{SIR_lrg_dir}
\end{figure}

Because of the ill-conditioning of $\Sigma$ (about $3\cdot 10^5$), we chose to perform a PCA first on the data, in order to improve the conditioning. 
SIR is ran on the data projected on the subspace of dimension 6 given by PCA, so that we can see the directions given by SIR on Fig. \ref{SIR_dep_dir} and Fig. \ref{SIR_lrg_dir}. The choice of the dimension of the subspace given by PCA as a preprocessing will be discussed in section 5.

We can see on Fig. \ref{SIR_dep_dir} that most of the information about the depression is held by the first 2 directions (top-right and top-center) given by SIR. Indeed we can get an accurate estimator of the depressions from the projection of the data on the second direction (top-center of Fig. \ref{SIR_dep_dir}) as these two quantities are linearly related.

We show on Fig. \ref{SIR_lrg_dir}, that the direction which is the most relevant to explain the width from the data highly depends on the value of the projected data on the several directions. For instance the first direction is relevant for high values of $x^Tu_1$ as the associated values of $y$ are narrow (top-left of Fig. \ref{SIR_lrg_dir}). The same first direction though is not very relevant to infer $y$ when the value of $x^Tu_1$ is around 0 because the corresponding values of $y$ spread from 1.5 to 3 which is 3/4 of the total range of values $y$ can take.

From Fig. \ref{SIR_dep_dir} and Fig. \ref{SIR_lrg_dir}, one can see that the directions which are relevant for estimating $y$ can vary from an individual to another.
So the question remains: which of these directions will be relevant to estimate $y$ from $x$?
We will suggest a way in the following section to individually select the most relevant sub-space for each of the elements $x$ one wants to determine the paired $y$.

\subsection{Direction selection}

As shown in Fig. \ref{SIR_dep_dir} and \ref{SIR_lrg_dir}, the directions computed by SIR that best explain $y$ from $x$ may depend on the value of $x$. This is why we suggest to select the directions that will minimize the variance on the values of $y$ associated with the $x^Tv$ which are in the neighbourhood of the $x_i$ one wants to determine the associated value $\hat{y}_i$. 
We thus can have a variable number of directions that is optimal to determine $\hat{y}_i$ from an individual to another.
For every individual $x_i$, we look for the set of directions $D_i$ spanning the subspace which minimizes the variance of $y_0 \in W_i$, where $W_i = \{y_{0_j} | dists_{i_j} \leq dist_{10_i}\}$, $dist_{i_j} = \| x_i^TD_i - x_j^TD_i \|$ and $ dist_{10_i}$ is the distance between $x_i^TD_i$ and its $10^{th}$ nearest neighbour.

\subsection{Validation of the modified SIR method}

The first test we will run is a comparison of the results given with the first direction of PCA, the first direction of SIR, and one selected direction of SIR.

\begin{figure}[H]

\hspace{-0.5cm}
\includegraphics[height = 5cm,width =10cm]{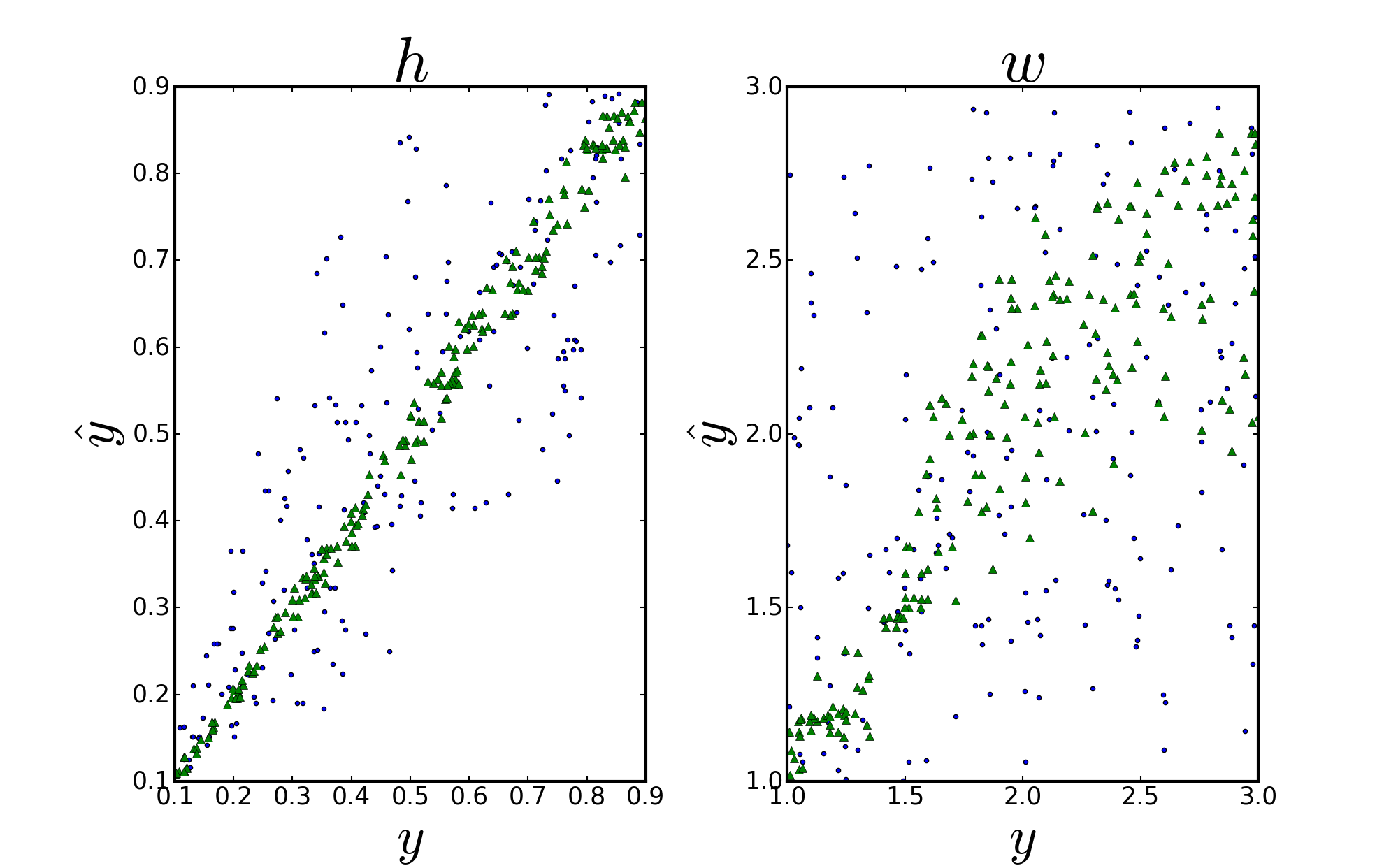}
\caption{Results of the estimation of depressions and widths with PCA using only the first direction (blue spots), SIR using the first direction (green triangles).}
\label{comp1dir}
\end{figure}

We can observe in Fig. \ref{comp1dir} that the results given by SIR are much more precise than the ones given by PCA. The perfect estimation being the line $\hat{y}=y$, results given by the method using SIR are closer to that line than the PCA ones.

\begin{figure}[H]
\hspace{-0.5cm}
\includegraphics[height = 5cm,width =10cm]{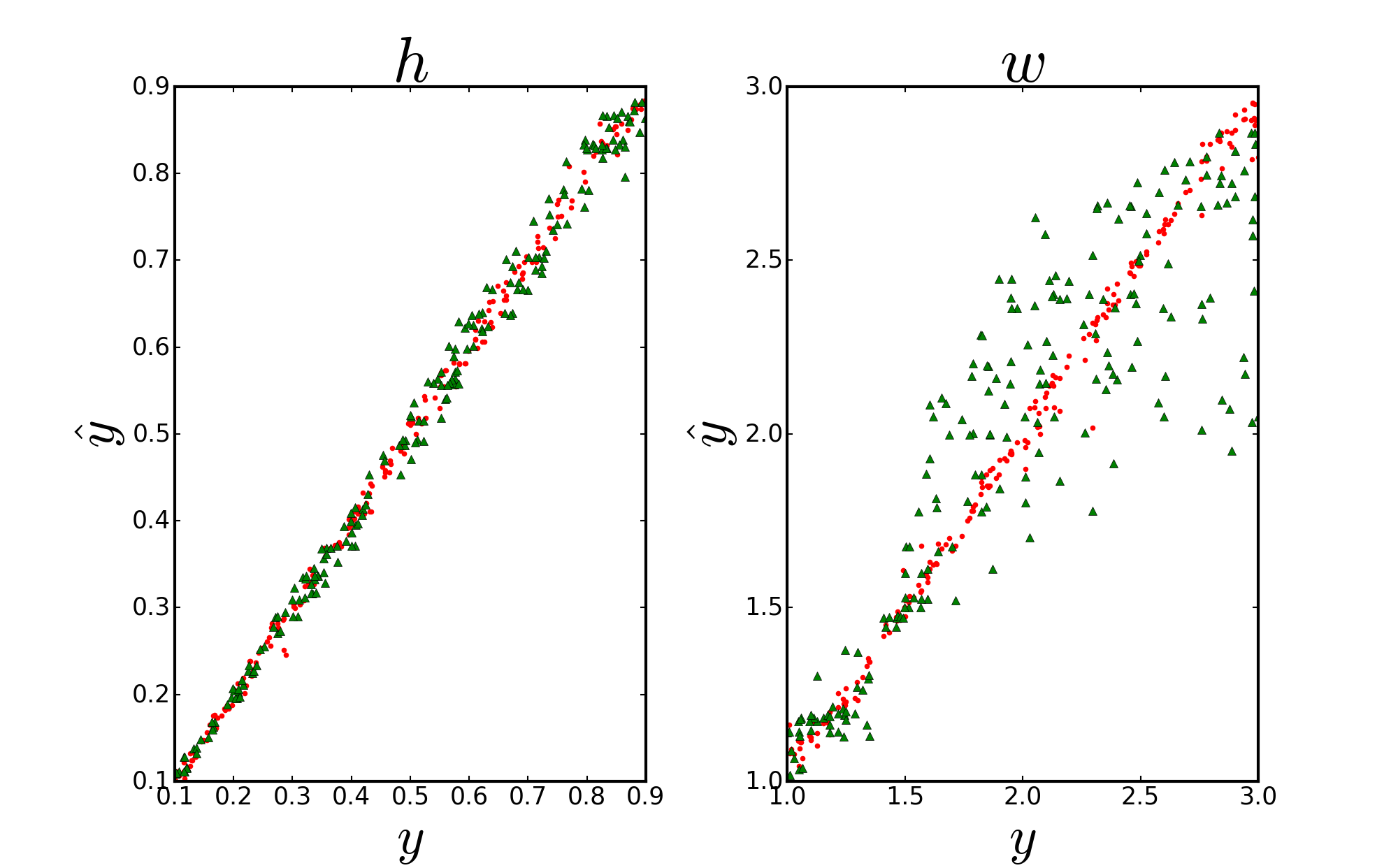}
\caption{Results of the estimation of depressions and widths with SIR using the first direction (green triangles), and the modified version of SIR with the direction selection (red spots).}
\label{comp1dir2}
\end{figure}

As shown in Fig. \ref{comp1dir2}, the selection of the direction that minimizes the variance of $y_0$ in the neighbourhood improves the results of SIR, and the efficiency of these different methods can be assessed with the computation of the mean absolute error (MAE) defined as:
\begin{equation}\label{eq_ame}
MAE = \frac{1}{M} \sum_{i=1}^M \| \hat{y}_i -y_i \|
\end{equation}  

\begin{table}[H]
\begin{tabular}{|c|p{2cm}|p{2cm}|p{2cm}|}
\hline
& Results with PCA first direction & Results with SIR first direction & Results with SIR one selected direction \\
\hline
Depressions &  0.097 & 0.012 & 0.010 \\
\hline
Widths &  0.616 & 0.223 & 0.029 \\
\hline
\end{tabular}
\vspace{.1cm}
\caption{\textnormal{Table of the mean absolute errors given by the three versions of the method using only one direction for the projection.}}
\label{table1}
\end{table}

The improvement given by the selection of the directions is very efficient in finding the best direction when the link between $x$ and $y$ is non linear.

We thus can compare the results given by the optimal version of PCA (keeping 6 components) with SIR on its optimal number of directions for either depression (2 directions) or widths (3 directions), and also the version of SIR which looks for the subspace which minimizes the variance on $y$ in the neighbourhood. The results of this comparison are shown on table \ref{table2}.

\begin{table}[H]
\centering
\begin{tabular}{|c|p{2cm}|p{2cm}|p{2cm}|}
\hline
 & PCA & Classic SIR \\
\hline
Depressions &  0.0063 & 0.0025 \\
\hline
Widths &  0.029 & 0.027 \\
\hline
\end{tabular}
\vspace{0.1cm}
\caption{\textnormal{Table of the mean absolute errors given by PCA with 6 directions, classic SIR with the optimal number of directions, and SIR with the selection of the directions which minimize the variance on $y_0$ in the neighbourhood of the individual one wants to estimate the value of $y$.}}
\label{table2}
\end{table}
\vspace{-0.8cm}

\begin{figure}[H]
\centering
\includegraphics[width =9cm,height = 4.5cm]{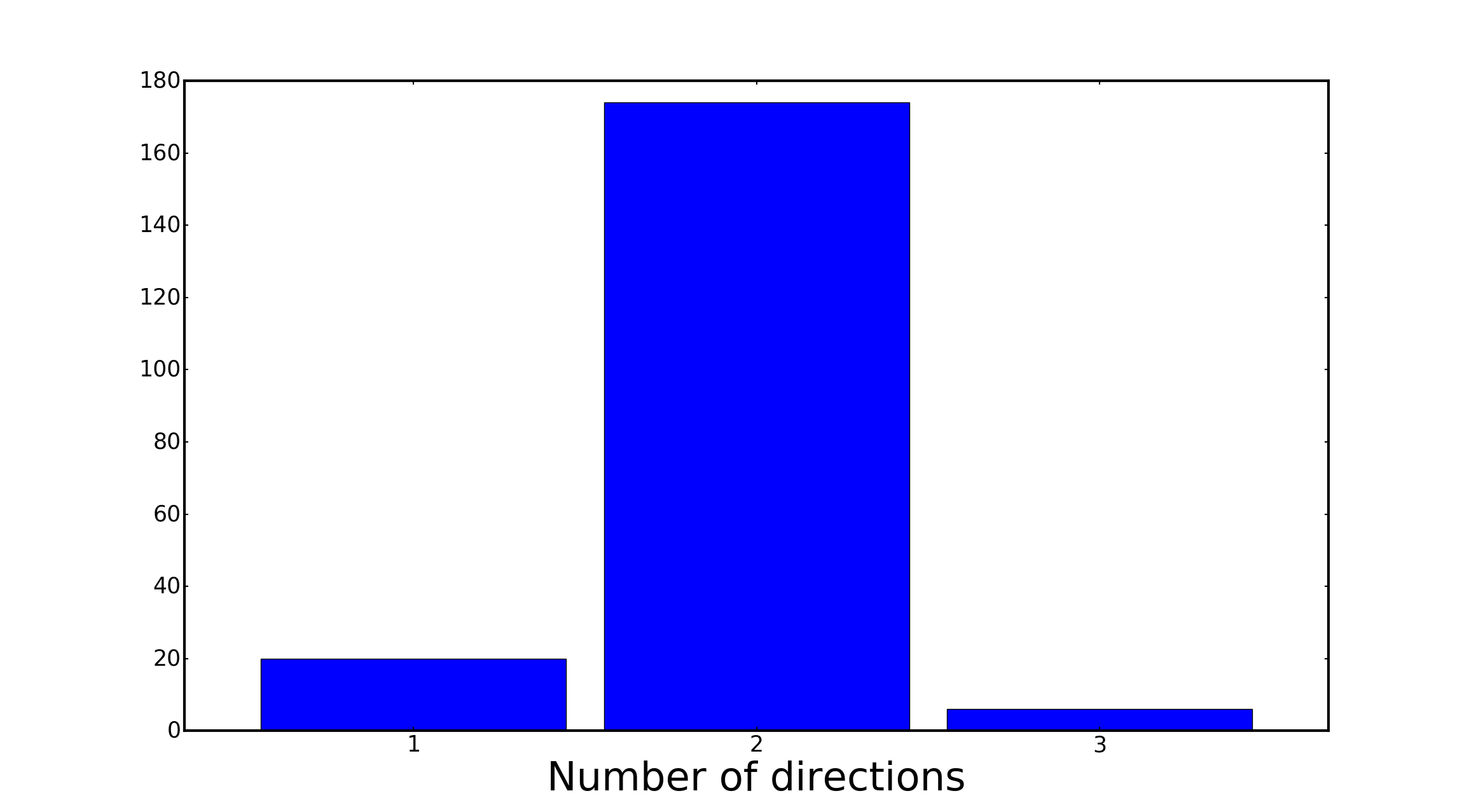}
\caption{Histogram of the dimension of the subspaces chosen by the method for the depression estimation for each of the 200 elements of the test database.}
\label{histoh}
\end{figure}

\begin{figure}[H]
\centering
\includegraphics[width =9cm,height = 4.5cm]{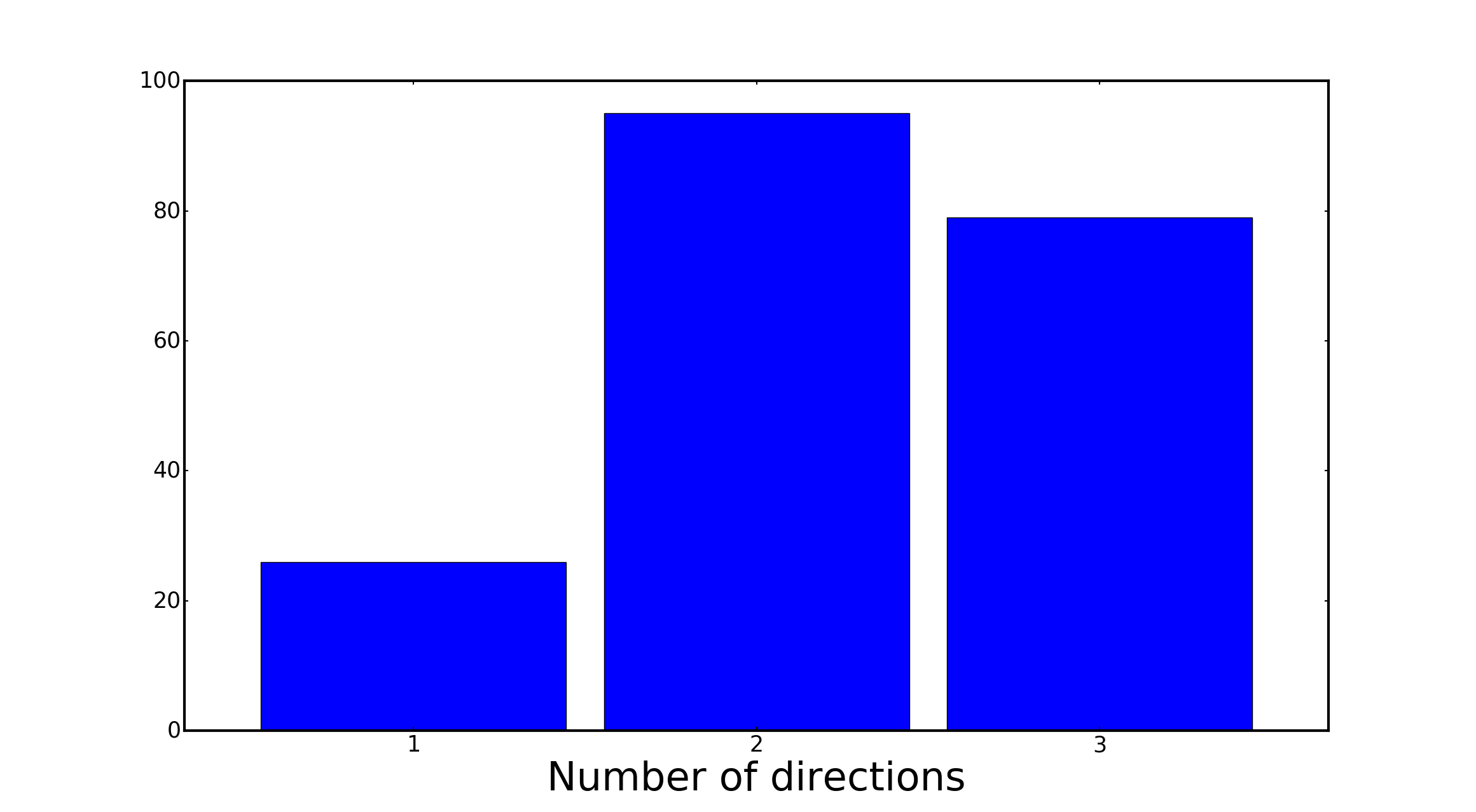}
\caption{Histogram of the dimension of the subspaces chosen by the method for the widths estimation for each of the 200 elements of the test database.}
\label{histow}
\end{figure}

The results displayed on table \ref{table2} show that SIR provides better results than the ones we can achieve with PCA. The fact that the classical version of SIR yields accurate estimates regarding the depressions can be explained by its efficiency on linear relationships between $x$ and $y$. Maybe the criterion used to determine the best directions combination (variance on $y$) is not the best criterion. However there is a non negligible gain regarding the widths estimation with SIR combined with the direction selection. We can note that the dimension of the subspace varies from an individual to another as shown on Fig. \ref{histoh} for the depressions and Fig. \ref{histow} for the widths.

We can observe in Fig. \ref{histoh} that, as in the classical SIR case, most of the time the subspace that best explains the depression value from the data is spanned by 2 directions. 

\vspace{-0.3cm}
Fig. \ref{histow} shows that the optimal number of directions to determine the width can either be 2 or 3. With classical SIR the optimal number of directions is 3, but Fig. \ref{histow} shows that it is not that often the best solution.

 \vspace{-0cm}
\section{Discussion about preprocessing}\label{seq_Test}

We chose to perform a PCA to improve the conditioning of $\Sigma$, as it is a way to reduce the dimension of the data-space, while loosing the least information. Doing that, we assume that the information held by the last directions given by PCA are, even if relevant, drawn into noise. Performing a PCA carefully should not cause any loss of information, except for the information we could not have retrieved anyway. But how many directions should be kept? If we keep too many, we will not make the conditioning decrease enough, and the directions given by SIR will not be accurate. If we do not keep enough directions, there is a risk of loosing information on the explanatory variable.

To get an idea of what the best solution would be, let us have a look at Fig. \ref{PCA_dep_dir} and Fig. \ref{PCA_lrg_dir} to the 10 first directions given by PCA on our dataset:

\begin{figure}[H]
\centering
\includegraphics[height = 5cm,width =9cm]{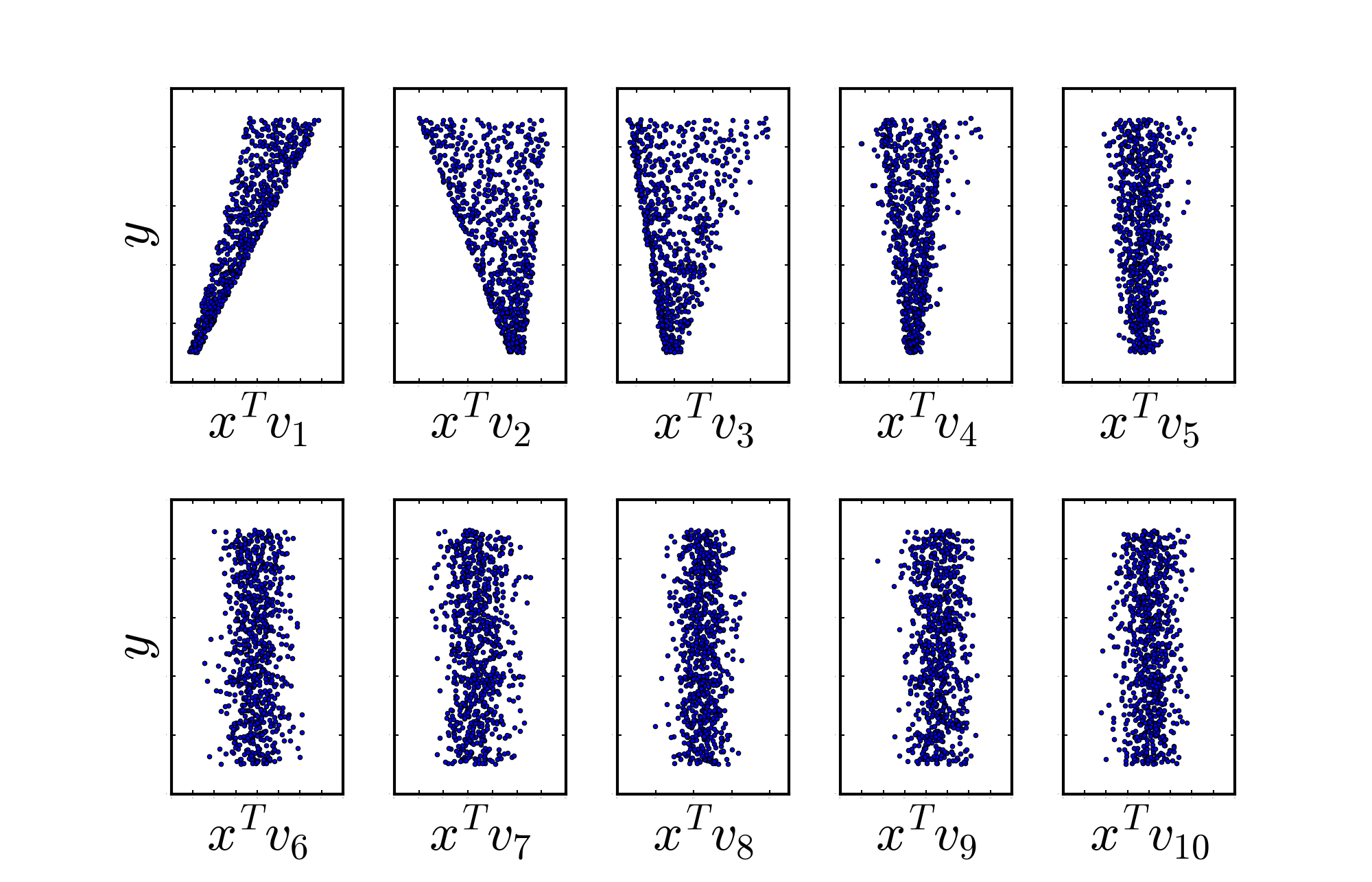}
\caption{Value of the depressions of the Gaussians $y$ versus the realisations of $\textbf{X}_0$ projected on the 10 directions given by PCA. Directions given by PCA seem to enclose information about $y$ up to the fifth direction.}
\label{PCA_dep_dir}
\end{figure}

\begin{figure}[H]
\centering
\includegraphics[height = 5cm,width =9cm]{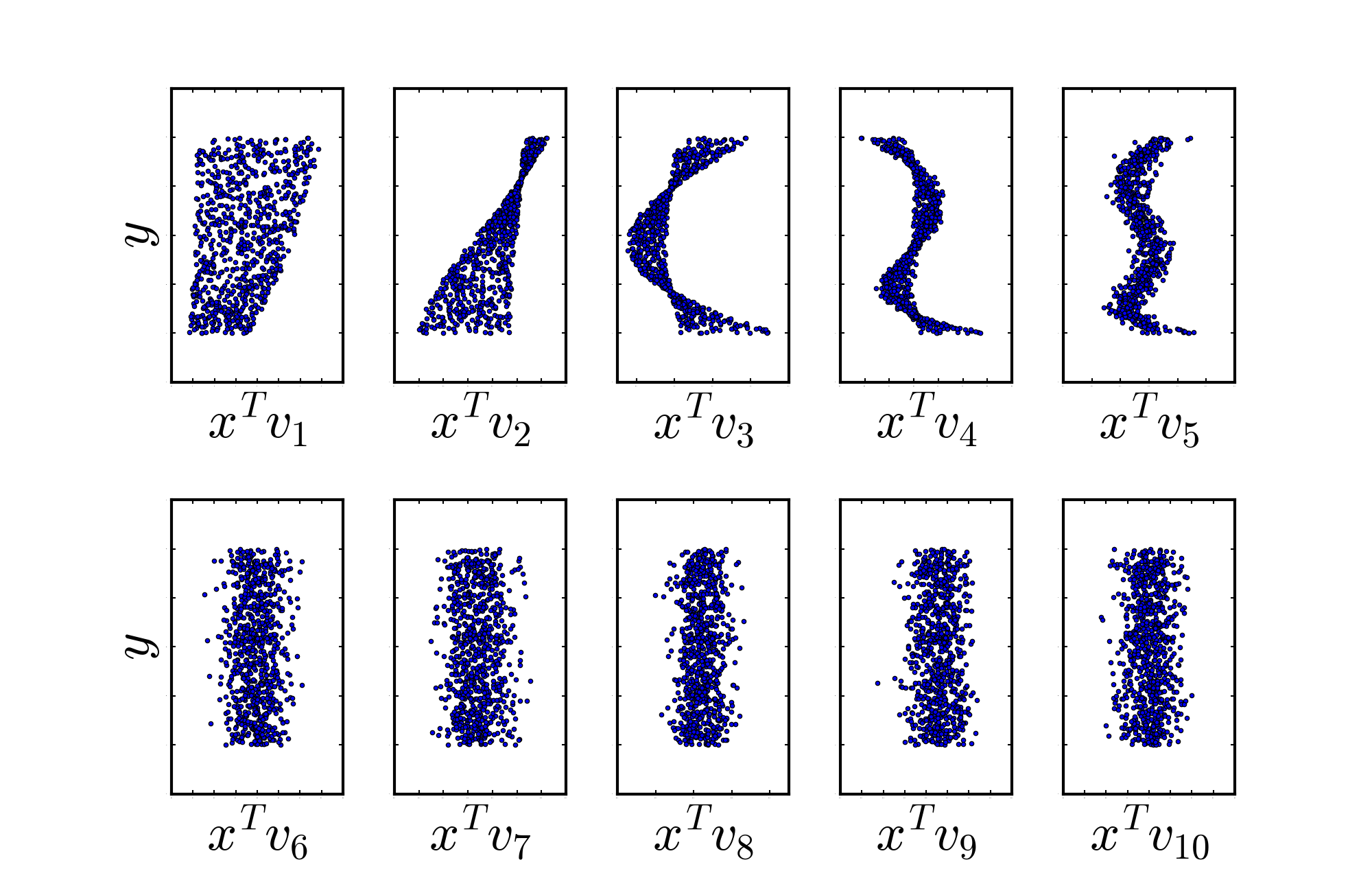}
\caption{Value of the widths of the Gaussians $y$ versus the realisations of $\textbf{X}_0$ projected on the 10 directions given by PCA. As in Fig. \ref{PCA_dep_dir}, the directions over the fifth seems to be irrelevant as the distribution of $y$ looks independent from the value of $x^Tv_i$.}
\label{PCA_lrg_dir}
\end{figure}

We can see in Fig. \ref{PCA_dep_dir} and Fig. \ref{PCA_lrg_dir} that from the sixth direction (bottom-left), the remaining directions do not seem to hold any information about the parameter, since the value of $y$ looks independent from $x^Tv_i$. This assumption is verified by a validation protocol based on the estimation of the explanatory variables with PCA for several noise realisations. It appears that the dimension of the subspace given by PCA that gives the best results is 6. It makes the conditioning decrease to about $4 \cdot 10^4$. Even if we only gain a 10 factor on the conditioning, we cannot reduce more the dimensionality because we would most probably loose important information about the explanatory variable.

\section{Conclusion and future work}\label{seq_ccl}

In this article we have shown that SIR, being a supervised method, provides a subspace more relevant to link the data to the explanatory variable than the PCA one. We have also tried to give an answer to the problem of the number of directions which would be relevant. We also considered the problem of the choice of that relevant directions which can change with the location of the studied sample. We have shown that the method can be adapted to give the optimal subspace to locally contain the manifold where the explanatory variable is best explained. Preliminary results for this method using the same data set as in \cite{Paletou15}, show that, compared to the PCA-based method, the proposed SIR-based method decreases the estimation error by $20\%$ for the effective temperature and the surface gravity while we obtain similar estimation errors for the metallicity. However, the error increases by $30\%$ for the projected rational velocity.
The adaptation of the method to real astrophysical data is still in progress. Because the data are a lot more complex, the dimension of the original data-space can be 10 or 100 times larger than the one used for this study, and the conditioning of $\boldsymbol{\Sigma}$ can reach $10^{23}$. We will need to run more tests to determine the optimal number of components to keep with PCA as a preprocessing, and also to know the best way to consider the neighbourhood depending on the number of directions kept by the method. We believe that better results should be achieved by finding a way to correctly tune the method for this particular case. 

\bibliographystyle{IEEEtran}
\bibliography{IEEEabrv,References}

\end{document}